\documentclass{article}

\usepackage{PRIMEarxiv}

\usepackage[utf8]{inputenc} 
\usepackage[T1]{fontenc}    
\usepackage{hyperref}       
\usepackage{url}            
\usepackage{booktabs}       
\usepackage{amsfonts}       
\usepackage{nicefrac}       
\usepackage{microtype}      
\usepackage{lipsum}
\usepackage{fancyhdr}       
\usepackage{graphicx}       
\graphicspath{{media/}}     

\usepackage{amsfonts,amsmath,amssymb,bm}

\usepackage[per-mode=fraction,exponent-product=\cdot]{siunitx}
\usepackage{booktabs, xcolor}
\usepackage{tabularx}
\usepackage{hyperref}
\hypersetup{%
  colorlinks=false,
  linkbordercolor=black,
  pdfborderstyle={/S/U/W 1}
}
\newcolumntype{C}{>{\centering\arraybackslash}X}

\usepackage{tikz}
\usetikzlibrary{positioning, arrows.meta, shapes.geometric}

\usepackage{subcaption}

\title{METAMAT 01: A semi-analytic Solution for Benchmarking Wave Propagation Simulations of homogeneous Absorbers in 1D/3D and 2D } 

\author{ \href{https://orcid.org/0000-0002-2148-6703}{\includegraphics[scale=0.06]{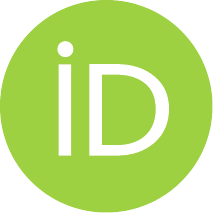}\hspace{1mm}Stefan~Schoder}\\
	TU Graz, IGTE-AVG\\
	8010 Graz, Austria \\
	\texttt{stefan.schoder@tugraz.at} \\
	 \AND
	 Paul Maurerlehner \\
  	TU Graz, IGTE-AVG\\
	8010 Graz, Austria \\
}

\newcommand{\eqVar}[1] {#1_{\mathrm{eq}}}
\newcommand{\acVar}[1] {#1^{\mathrm{ac}}}

\newcommand{\idxa}{j}
\newcommand{\idxb}{k}
\newcommand{\idxc}{l}
\newcommand{\idxd}{m}

\newcommand\etal{\textit{et al. }}
\newcommand{\tenPot}[1] {$\times10^{{#1}}$}

\begin{document}
\maketitle

\begin{abstract}
The development of acoustic simulation workflows in the time-domain description is essential for predicting the sound of aeroacoustic \cite{schoder2019hybrid,schoder2022aeroacoustic,schoder2023acoustic} or other transient acoustic effects. A common practice for noise mitigation is using absorbers. The modeling of these acoustic absorbers is typically provided in the frequency domain. Several, methods established bridging this gap, investigating methods to model absorber in the time domain. Therefore, this short article, describes the analytic solution in time-domain for benchmarking absorber simulations with infinite 1D, 2D, and 3D domains. Connected to the analytic solution, a \textit{Matlab} script is provided to easily obtain the reference solution. The reference codes are provided as benchmark solution in the EAA TCCA Benchmarking database as \href{https://zenodo.org/communities/eaa-computationalacoustics}{METAMAT 01} \cite{Schoder-METAMAT}.
\end{abstract}


\section{Acoustic meta-material, effective parameters and dispersion relation}
Applying the inverse Fourier transform to the Helmholtz equation for a frequency-dependent equivalent bulk modulus $\eqVar{K}(\omega)$ and density $\eqVar{\rho}(\omega)$ yields in time domain
\begin{equation}
\label{eq:inhomogWaveEq_time}
\{\frac{1}{\eqVar{K}} * \acVar{\ddot{p}}\}(t)
- \{\nabla \cdot \left( \frac{1}{\eqVar{\rho}} * \nabla \acVar{p} \right)\}(t)  = 0 \, ,
\end{equation}
with $\acVar{p}$ denoting the acoustic pressure, $\{a*b\}(t) = \int_{-\infty}^{\infty} a(t-\tau)\cdot b(\tau)\ d\tau$ the time-convolution operator, and the two dots above the variable indicating the second-order partial time derivative. The dispersion relation of the frequency-dependent material reads as
\begin{equation}
    |\boldsymbol{k}|^2=\omega^2 \eqVar{\hat \rho}(\omega) \eqVar{\hat K}^{-1}(\omega) .
\end{equation}
The solution of \ref{eq:inhomogWaveEq_time} depends only on the norm $|\boldsymbol{k}|^2=\boldsymbol{k}^{\top} \boldsymbol{k}$ of the wavevector and this is a consequence of the isotropy of the medium. The phase velocity $\bm{v}_{\mathrm{p}}$ and the group velocity $\bm{v}_{\mathrm{g}}$ are then defined by
$$
\bm{v}_{\mathrm{p}}(\boldsymbol{k})=\omega(|\boldsymbol{k}|) \frac{\boldsymbol{k}}{|\boldsymbol{k}|^2} \quad \text { and } \quad \bm{v}_{\mathrm{g}}(\boldsymbol{k})=\nabla_{\boldsymbol{k}} \omega(|\boldsymbol{k}|) \, ,
$$
with $\nabla_{\boldsymbol{k}}$ the gradient regarding $\boldsymbol{k}$.

\section{Analytic solution in 1D}

Bellis and Lombard \cite{bellis2019simulating} presented a reference semi-analytical solution for 1D wave propagation for acoustic metamaterial using auxiliary fields by applying time and space Fourier transforms to the governing equations. A slight modification of this solution is presented here for a semi-infinite duct $x\in [0,+\infty [$ with plane waves (allowing a 1D representation). The waves are excited based on a Dirichlet boundary condition at $x=x_0=0$, of arbitrary shape. The Fourier transform $\hat{p}$ of the acoustic pressure $p$ reads
\begin{equation}
    \hat{p}(x, \omega)=e^{-i k(\omega)\left|x-x_s\right|} \hat{p_0}(\omega)\, ,
    \label{eq:analSol}
\end{equation}
with $k(\omega)$ being the wave number that satisfies the dispersion relation, $\omega$ the angular frequency, $x$ the space coordinate, $x_0$ the location of the boundary with inhomogenous Dirichlet boundary condition with the Fourier transform of the transient excitation signal $\hat{p}_0(\omega)$. 

\section{Analytic solution in 2D}
For a 2D infinite disc with a hole at $r_0$, i.e., $r\in [r_0,+\infty [$ with an inhomogeneous Dirichlet boundary condition at $r=r_0$, the Fourier transform $\hat{p}$ of the acoustic pressure $p$ reads  
\begin{equation}
    \hat{p}(x, \omega)=A(\omega) H_0^{(2)}(k(\omega) r) \, ,
\end{equation}
where $A(\omega) = \hat{p_0}(\omega)/  H_0^{(2)}(k(\omega) r_0)$ is the adapted coefficients for satisfying the boundary condition at $r=r_0$ in time domain (see respective Matlab script) and $H_0^{(2)}(k(\omega) r)$ denotes the zeroth Hankel function of second kind.

\section{Analytic solution in 3D}
Based on the inherent nature of the wave equation, the 3D analytic solution (infinite domain with a spherical hole with radius $r_0$., i.e., $r\in [r_0,+\infty[$) can be based on the 1D solution with a $1/r$ scaling and the boundary condition adjustment as done in the 2D case for $r=r_0$.





\section{Benchmark example of a passive absorber 1D and 2D}

\subsection{Transient excitation $p_0(t)$}
Boundary condition function is defined as follows:
\begin{equation}
    p_0(t)=\left\{\begin{array}{l}
\sum_{m=1}^4 a_m \sin \left(\beta_m \omega_c t\right) \quad \text { if } 0<t<\frac{1}{f_c}, \\
0 \text { otherwise, }
\end{array}\right.
\end{equation}
Setting the coefficients to $\beta_m=2^{m-1}$, $a_1=1, a_2=-21 / 32, a_3=63 / 768,$ and $ a_4=-1 / 512$ entails the smoothness property $p_0 \in C^6([0,+\infty[)$. In Fig.~\ref{fig:transDuct_excSig}, the signal is shown for a central frequency $f_c=\omega_c / 2 \pi=700 \mathrm{~Hz}$.

\subsection{Material}

\begin{figure}[htb]
	\centering
    \includegraphics[height=3.95cm]{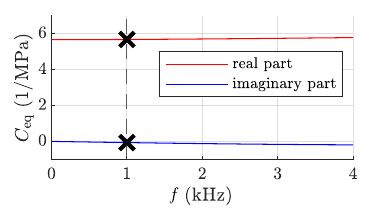}
    \hspace{0.5cm}
    \includegraphics[height=3.95cm]{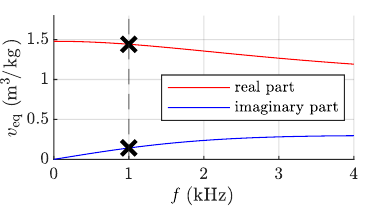}
    \caption{Real and imaginary parts of the equivalent compressibility $\eqVar{\hat C}(f)$ (left) and equivalent specific volume $\eqVar{\hat v}(f)$ (right) as a function of frequency for the fictitious porous material \textit{mat1}. }
    \label{fig:frfs_strongPorous}
\end{figure}

For the analysis, the fictitious porous material \textit{mat1} derived from  Yoshida \etal \cite{yoshida2020TDEF} is considered. The EF model exhibits strong damping. The FRFs of the equivalent compressibility $\eqVar{\hat C}(f)$ and specific volume $\eqVar{\hat v}(f)$ are depicted in Fig.~\ref{fig:frfs_strongPorous}.
The corresponding rational function approximation coefficients of the EF parameters according to the rational function definition \eqref{eq:multiPoleModel_C} and \eqref{eq:multiPoleModel_v} are listed in Tab.~\ref{tab:FRF_poles_v_mat1Strong}.
\begin{table}[htb]
    \begin{subtable}[h]{0.45\textwidth}
    \centering
    \caption{Equivalent compressibility $\eqVar{\hat C}(\omega)$.}
    \vspace{2mm}
    \label{tab:FRF_poles_C_mat1Strong}
	\begin{tabular}{l r r}
	\toprule
	$j$    &$A^{C}_j$ ($\frac{\mathrm s}{\mathrm{Pa}}$) & $\alpha^{C}_j$ ($\frac{1}{\mathrm s}$) \\ \midrule
	1    &3.32994\tenPot{-1}       & 591390           \\
	2    &1.37113\tenPot{-1}        & 37500           \\
	$C_{\infty}$ 	& 6.5866\tenPot{-6}      & $-$               \\
 \bottomrule
	\end{tabular}
    \end{subtable}
    \hfill
    \begin{subtable}[h]{0.45\textwidth}
    \centering
    \caption{Equivalent specific volume $\eqVar{\hat v}(\omega)$.}
    \vspace{2mm}
    \label{tab:FRF_poles_v_mat1Strong}
	\begin{tabular}{l r r}
	\toprule
	$l$       &$A^{v}_l$ ($\frac{\mathrm{m}^3}{\mathrm{kg}\, s}$) & $\alpha^{v}_l$ ($\frac{1}{\mathrm s}$) \\ \midrule
	1       &$-$20397.84         & 84764   \\
	2       &$-$10110.46          & 21687   \\ 
	$v_{\infty}$ & 0.772093         &  $-$                 \\
\bottomrule                       
	\end{tabular}
\end{subtable}
\caption{Coefficients of the rational function approximation of the equivalent parameters of the fictitious material \textit{mat1}: residues $A^{ }$, poles $\alpha^{ }$, and constants $C_{\infty}$ and $v_{\infty}$.}
\label{tab:FRF_poles_mat1Strong}
\end{table}
%
In doing so, the FRF of the equivalent compressibility $\eqVar{\hat C}(\omega)$ being the inverse of the equivalent compression modulus ${\eqVar{\hat K}}$ and the equivalent specific volume $\eqVar{\hat v}(\omega)$ being the inverse of the equivalent density ${\eqVar{\hat \rho}}$ are approximated by a sum of rational functions
\begin{subequations}
\begin{align}
\label{eq:multiPoleModel_C}
\eqVar{\hat C}(\omega)&=\frac{1}{\eqVar{\hat K}} = 
C_{\infty} 
+ \sum_{\idxa=1}^{N^{C}} \frac{A_\idxa^{C}}{\alpha_\idxa^{C}-i \omega} 
+ \frac{1}{2} \sum_{\idxb=1}^{M^{C}} \left[\frac{B_\idxb^{C}+i C^{C}_\idxb}{\beta_\idxb^{C}+i \gamma_\idxb^{C}-i \omega}
+\frac{B_\idxb^{C}-i C_\idxb^{C}}{\beta_\idxb^{C} - i \gamma_\idxb^{C}-i \omega}\right] \, \\
\label{eq:multiPoleModel_v}
\eqVar{\hat v}(\omega)&=\frac{1}{\eqVar{\hat \rho}} = 
v_{\infty} 
+ \sum_{\idxc=1}^{N^{v}} \frac{A_\idxc^{v}}{\alpha_\idxc^{v}-i \omega} 
+ \frac{1}{2} \sum_{\idxd=1}^{M^{v}} \left[\frac{B_\idxd^{v}+i C_\idxd^{v}}{\beta_\idxd^{v} + i \gamma_\idxd^{v}-i \omega}
+\frac{B_\idxd^{v}-i C_\idxd^{v}}{\beta_\idxd^{v}-i \gamma_\idxd^{v}-i \omega}\right] \, .
\end{align}
\end{subequations}
Herein, the superscripts $C$ and $v$ indicate the assignment of coefficients and variables to the equivalent compressibility $\eqVar{C}$ and specific volume $\eqVar{v}$ and $i$ denotes the imaginary unit.
The order of the approximation is defined by the number of real poles $N^{C}$ and $N^{v}$ and complex conjugate pole pairs $M^{C}$ and $M^{v}$, respectively.

\subsection{Matlab code}
The reference Matlab code (1D and 2D) to plot the semi-analytical solutions is hosted in the following Zenodo repository of the EAA TCCA Benchmarking database.

\subsection{Finite element simulation using openCFS}
The reference openCFS \cite{schoder2022opencfs} files for a simulation run is hosted in the following Zenodo repository of the EAA TCCA Benchmarking database.

\subsection{Results in 1D}

\begin{figure}[htb]
     \centering
     \begin{subfigure}[b]{0.38\textwidth}
     \centering
         \includegraphics[height=5.6cm]{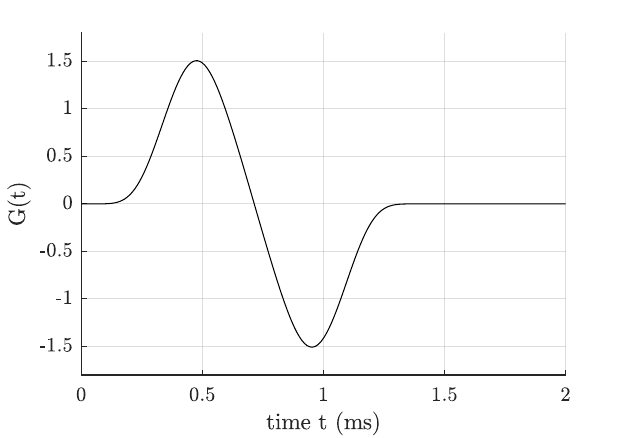}
         \caption{Transient excitation signal.}
         \label{fig:transDuct_excSig}
     \end{subfigure}
     \hspace{-0.2cm}
     \begin{subfigure}[b]{0.6\textwidth}
     \centering
         \includegraphics[height=5.6cm]{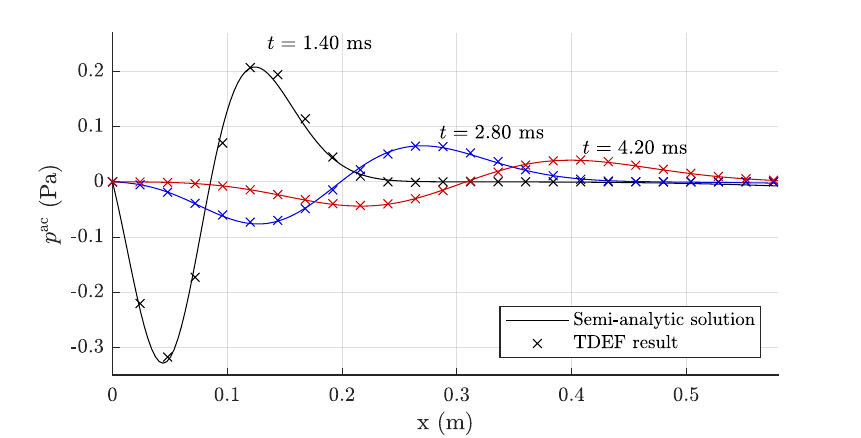}
         \caption{Results at selected time steps.}
         \label{fig:transDuct_res}
     \end{subfigure}
     \caption{Comparison of the semi-analytic solution (according to \eqref{eq:analSol}) and the FE result based on the TDEF formulation (labelled as "TDEF result").}
     \label{fig:transDuct}
\end{figure}

The analytical solution of the 1D problem is given by \eqref{eq:analSol}. By computing the inverse FFT of \eqref{eq:analSol} and taking the real part, the solution of \eqref{eq:analSol} can be compared to the FE simulation result.
Figure~\ref{fig:transDuct_res} shows the semi-analytical and the TDEF-based numerical solution at three selected time steps.

\subsection{Results in 2D}

The FEM simulation results on the 2D example can be found in \cite{Maurerlehner2024}.

\section{Conclusion} 
In conclusion, this short article addresses the need for the development of acoustic simulation workflows in the time-domain, which can be beneficial in terms of computational effort (e.g. direct computation of impulse responses in room acoustics \cite{refId0}) or avoid the transformation (FFT) of  aeroacoustic phenomena based on unsteady flow simulations (like tissue in phonation \cite{dollinger2023overview,kraxberger2023alignment,schoder2020hybrid}, confined flow aeroacoustics \cite{maurerlehner2022aeroacoustic}, or fan noise \cite{schoder2020computational,tieghi2023machine,schoder2023affordable,antoniou2023numerical}). The dispersive nature of sound-absorptive material (e.g., foam) or acoustic metamaterials entails time convolution integrals in time-domain formulations of the governing equations. Various formulations have been established, to solve these convolutions efficiently, where either additional degrees of freedom are introduced or part of the history of the solution variable needs to be stored. Thereby, the goal is to keep the additional computational effort to a minimum. We contribute to the ongoing efforts by presenting analytic solutions in the time domain, specifically designed for benchmarking absorber simulations. The inclusion of a \textit{Matlab} script facilitates the straightforward acquisition of the reference solution, offering a practical tool for researchers in the field. The provided reference codes are made publicly accessible as a benchmark solution in the EAA TCCA Benchmarking database under the identifier \href{https://zenodo.org/communities/eaa-computationalacoustics}{METAMAT 01}.


\bibliographystyle{unsrt}  
\bibliography{references}

\begin{thebibliography}{10}

\bibitem{schoder2019hybrid}
Stefan Schoder and Manfred Kaltenbacher.
\newblock Hybrid aeroacoustic computations: State of art and new achievements.
\newblock {\em Journal of Theoretical and Computational Acoustics}, 27(04):1950020, 2019.

\bibitem{schoder2022aeroacoustic}
Stefan Schoder, Manfred Kaltenbacher, {\'E}tienne Spieser, Hugo Vincent, Christophe Bogey, and Christophe Bailly.
\newblock Aeroacoustic wave equation based on pierce's operator applied to the sound generated by a mixing layer.
\newblock In {\em 28th AIAA/CEAS Aeroacoustics 2022 Conference}, page 2896, 2022.

\bibitem{schoder2023acoustic}
Stefan Schoder, {\'E}tienne Spieser, Hugo Vincent, Christophe Bogey, and Christophe Bailly.
\newblock Acoustic modeling using the aeroacoustic wave equation based on pierce’s operator.
\newblock {\em AIAA Journal}, pages 1--10, 2023.

\bibitem{Schoder-METAMAT}
Stefan Schoder and Paul Maurerlehner.
\newblock {METAMAT 01: A Semi-analytic Solution for Benchmarking Wave Propagation Simulations of homogeneous Absorbers}, March 2024.

\bibitem{bellis2019simulating}
C{\'e}dric Bellis and Bruno Lombard.
\newblock Simulating transient wave phenomena in acoustic metamaterials using auxiliary fields.
\newblock {\em Wave Motion}, 86:175--194, 2019.

\bibitem{yoshida2020TDEF}
Takumi Yoshida, Takeshi Okuzono, and Kimihiro Sakagami.
\newblock Time-domain finite element formulation of porous sound absorbers based on an equivalent fluid model.
\newblock {\em Acoustical Science and Technology}, 41(6):837--840, 2020.

\bibitem{schoder2022opencfs}
S.~Schoder and K.~Roppert.
\newblock {openCFS}: Open source finite element software for coupled field simulation -- part acoustics, 2022.

\bibitem{Maurerlehner2024}
Paul Maurerlehner, Dominik Mayrhofer, Manfred Kaltenbacher, and Stefan Schoder.
\newblock A time-domain finite element formulation of the equivalent fluid model for the acoustic wave equation.
\newblock {\em Acta Acustica}, in review:--, 2024.

\bibitem{refId0}
{Kraxberger, Florian}, {Kurz, Eric}, {Weselak, Werner}, {Kubin, Gernot}, {Kaltenbacher, Manfred}, and {Schoder, Stefan}.
\newblock A validated finite element model for room acoustic treatments with edge absorbers.
\newblock {\em Acta Acust.}, 7:48, 2023.

\bibitem{dollinger2023overview}
Michael D{\"o}llinger, Zhaoyan Zhang, Stefan Schoder, Petr {\v{S}}idlof, Bogac Tur, and Stefan Kniesburges.
\newblock Overview on state-of-the-art numerical modeling of the phonation process.
\newblock {\em Acta Acustica}, 7:25, 2023.

\bibitem{kraxberger2023alignment}
Florian Kraxberger, Christoph N{\"a}ger, Marco Laudato, Elias Sundstr{\"o}m, Stefan Becker, Mihai Mihaescu, Stefan Kniesburges, and Stefan Schoder.
\newblock On the alignment of acoustic and coupled mechanic-acoustic eigenmodes in phonation by supraglottal duct variations.
\newblock {\em Bioengineering}, 10(12):1369, 2023.

\bibitem{schoder2020hybrid}
Stefan Schoder, Michael Weitz, Paul Maurerlehner, Alexander Hauser, Sebastian Falk, Stefan Kniesburges, Michael D{\"o}llinger, and Manfred Kaltenbacher.
\newblock Hybrid aeroacoustic approach for the efficient numerical simulation of human phonation.
\newblock {\em The Journal of the Acoustical Society of America}, 147(2):1179--1194, 2020.

\bibitem{maurerlehner2022aeroacoustic}
Paul Maurerlehner, Stefan Schoder, Johannes Tieber, Clemens Freidhager, Helfried Steiner, G{\"u}nter Brenn, Karl-Heinz Sch{\"a}fer, Andreas Ennemoser, and Manfred Kaltenbacher.
\newblock Aeroacoustic formulations for confined flows based on incompressible flow data.
\newblock {\em Acta Acustica}, 6:45, 2022.

\bibitem{schoder2020computational}
Stefan Schoder, Clemens Junger, and Manfred Kaltenbacher.
\newblock Computational aeroacoustics of the eaa benchmark case of an axial fan.
\newblock {\em Acta Acustica}, 4(5):22, 2020.

\bibitem{tieghi2023machine}
Lorenzo Tieghi, Stefan Becker, Alessandro Corsini, Giovanni Delibra, Stefan Schoder, and Felix Czwielong.
\newblock Machine-learning clustering methods applied to detection of noise sources in low-speed axial fan.
\newblock {\em Journal of Engineering for Gas Turbines and Power}, 145(3):031020, 2023.

\bibitem{schoder2023affordable}
Stefan Schoder, Jakob Schmidt, Andreas F{\"u}rlinger, Roppert Klaus, and Maurerlehner Paul.
\newblock An affordable acoustic measurement campaign for early prototyping applied to electric ducted fan units.
\newblock {\em Fluids}, 8(4):116, 2023.

\bibitem{antoniou2023numerical}
Evangelos Antoniou, Gianluca Romani, Andreas Jantzen, Felix Czwielong, and Stefan Schoder.
\newblock Numerical flow noise simulation of an axial fan with a lattice-boltzmann solver.
\newblock {\em Acta Acustica}, 7:65, 2023.

\end{thebibliography}

\end{document}